\documentclass[review]{elsarticle}
\usepackage{lineno,hyperref}
\usepackage{graphicx}
\usepackage{latexsym}
\usepackage{xcolor}
\usepackage{graphicx}
\usepackage{subfig}
\usepackage{amsmath}
\usepackage{amssymb}
\modulolinenumbers[5]
\journal{Journal of \LaTeX\ Templates}

\begin{document}
\newcommand{\eq}{\begin{equation}}                                                                         
\newcommand{\eqe}{\end{equation}} 
\begin{frontmatter}

\title{  Advanced self-similar solutions of regular and irregular diffusion equations  }

\author{Imre Ferenc Barna \fnref{myfootnote}}
\address{{Wigner Research Centre for Physics, Konkoly--Thege Mikl\'os \'ut 29--33}, Budapest, Hungary}
\ead[url]{www.kfki.hu/~barnai}
\cortext[mycorrespondingauthor]{Corresponding author}
\ead{barna.imre@wigner.hu}

\author{L\'aszl\'o M\'aty\'as} 
\address{Sapientia Hungarian University of Transylvania, Department of Bioengineering, 
Faculty of Economics, Socio-Human Sciences
and Engineering, Libert\u{a}tii sq. 1, Miercurea Ciuc, Romania}


\begin{abstract}
 We study the diffusion equation with an appropriate change of variables. This equation is in general a partial 
differential equation (PDE).    
With the self-similar and related Ansat\"atze we transform the PDE of diffusion to an ordinary differential equation. 
The solutions of the PDE belong to a family of functions which are presented for the case of infinite horizon.   
In the presentation we accentuate the physically reasonable solutions.  
We also study  time dependent diffusion phenomena, where the spreading may vary in time. 
To describe the process we consider time dependent diffusion coefficients.    
The obtained analytic solutions all can be expressed with Kummer's or Whittaker-type of functions.
\end{abstract}


\end{frontmatter}
%
\section{Introduction}
The study of transport processes are crucial both from scientific and from engineering points as well. 
One of such process is diffusion which is a quite general phenomena.  It can be formulated for particles 
which is the classical mass diffusion or to energy transport which it is called heat conduction.  
The recent manuscript is an extension of certain studies, in which we present additional analytic solutions 
for the regular diffusion equation from symmetry considerations. In our last paper \cite{barna1} after a historical overview 
we presented a class of analytic solutions derived with the help of the reduction mechanism 
based on different trial functions or Ans\"atze. As we will see this paper is second in that line giving more 
detailed analysis of various self-similar and generalized self-similar solutions. We will show how 
the original self-similar trial function can be generalized in a kind of power law expansion.   
  \\  
The whole diffusion phenomena is in general well introduced with cases and studies by the monograph of Ghez \cite{ghez}. 
Embedded into the larger field of transport processes diffusion is discussed by John Newman and 
Vincent Battaglia in their series of lectures \cite{neu}. 
Gillespie and Seitariodu gave an introduction to the standard theoretical models for simple Brownian type diffusion \cite{giles} in 2013. 
 The anomalous diffusion was analyzed with statistical methods by Weihua and co-workers \cite{weihua}. 
Uchaikin investigated the self-similar anomalous diffusion L\'evy-stable laws \cite{uch}.  
Ari Arapostathis {\it{et al.}} studied the ergodic control of diffusion processes in a monograph \cite{ari}.
Regarding applications in solid state physics, binary alloys, thin films 
etc.  one may find in \cite{SeAl2017,Dy2002,shev,adv,Ma90,BeEr2006,BaEr2008}. 
Defects and diffusion in nanotubes was summarized by Fisher \cite{fisher}. Diffusion processes a peculiar material in ceramics was analyzed 
in the monograph by Pelleg \cite{pelleg}.  Atomic diffusion processes is a  scientific field which was presented in a monograph as well \cite{atomstar}.  
Diffusion processes are the starting points for reaction diffusion processes \cite{liehr} or diffusions in porous media \cite{velaz}. 
A vitally important application of such mathematical equations is the  mathematical modeling of aircraft cabin fires \cite{fire,fire2}. 
The two dimensional diffusion equation, completed with certain reaction terms may lead to pattern formation, 
for instance the Turing patterns derived from Schnakenberg equations \cite{IrWeWi2004,KhZhFu21} or Brusselator model \cite{GoMaVo2008}. 
If the diffusion equation is completed with the simplest nonlinear term, the gradient of the variable on the second power we arrive to the 
Kardar-Parizi-Zwang (KPZ) equation which is the simplest successful model for surface growth phenomena. In two of our former studies we investigated 
the KPZ equation (with additional noise terms) with the self-similar \cite{imre_kpz1} and the traveling waves \cite{imre_kpz2} Ans\"atze.
   Interesting results has been obtained in the study of irregular diffusion  \cite{bbb1,bbb2}. 
Now, in the following we deepen this analysis and present additional analytic solutions with detailed parameter study.   
We present a series of solutions which are defined on the whole real axis having a decaying and spreading property with 
additional oscillations. Our main investigation tool, the self-similar Ansatz helps us to build a link from the regular to irregular diffusion processes 
which makes up the second part of the study. 
\section{Theory and Results}
Although the diffusion process can be studied in different dimensions, here we consider only 
 one Cartesian coordinate therefore the equation reads  
\begin{eqnarray}
\frac{\partial C(x,t)}{\partial t  } = D \frac{\partial^2 C(x,t)}{\partial x^2  }, 
\label{pde} 
\end{eqnarray}
where  $C(x,t)$ is the distributions of the particle concentration in space and time and $D$ 
is the diffusion coefficient. $C(x,t)$ in the equation above is considered up to a constant, consequently 
it may also refer to the concentration above or around the average. The 
function $C(x,t)$ fulfills the necessary smoothness conditions with existing continuous first and second derivatives in respect to time and and space. 
From causality the diffusion coefficient should be a positive real number  $(D > 0)$. Numerous 
physics textbooks gives us the derivation how the fundamental (the Gaussian) solutions can be obtained e.g. \cite{crank75,Bennett2013}.   

The derivation and analysis of various analytic solutions of physical processes described 
by various mathematical equations, (e.g. algebraic, differential or 
partial differential) have crucial importance. As it was shown in our former paper \cite{barna1} and as it will be shown here, there are 
far more solutions known for diffusion than the Gaussian and the error functions.   
We presented some trivial solutions (e.g.  $ t + Dx^2/2$), other solutions which can derived 
with the  general symmetry analysis method 
by Clarkson and Kruskal \cite{krusk}, the traveling profile method of Benhamidouche \cite{behn} or the self-similar Ansatz of Sedov \cite{sed}. 
Beyond the Gaussian and error functions most of our results are expressible with the Kummer's special functions. 
  In the following we try additional two trial functions and enlarge the number of solutions known from the self-similar Ansatz.  
In the last part of the study we investigate less regular diffusion processes where the diffusion coefficients gain temporal dependencies. 
The diffusion equation stands at the basis of more complex equations: in case on the r.h.s. beyond the second derivative, 
there is a function $F(C)$ with certain properties, 
we can talk about the Kolmogorov-Petrovskii-Piskunov equation \cite{KoPePi91}. Explicitly, on 
the r.h.s. the term  $C(1-C)$ yields the Fisher equation \cite{Fisher1937,Al-Khaled2001},  
the term $ p C + r C^q $ in general means the Newel-Whitehead-Segel (NWS) equation 
\cite{NeWh69,Se69} and  the term  $C(1-C)(C-\alpha)$  where $ 0 < \alpha < 1$ defines  
the general Zeldovich equation  (or Huxley equation) which arises in combustion theory \cite{zeld}.    
For certain non-trivial values of $p, r$ and $q$ one may find exact  solutions of the NWS equation \cite{GiKe2001,NoSoNa2011}.  
Frank-Kamenetzkii used the $\exp(a\cdot e^{-\frac{b}{C}})$ term \cite{kamen} to explain thermal explosion. 

Burgers used the $C\cdot C_x$ (where the subscript stands for partial derivation) term to study turbulence \cite{burgers}.
Nariboli and Lin introduced the quadratic Burgers equation 
with the term of $C^2\cdot C_x$ \cite{narib}. 
Sachdev  \cite{sachdev}  modified  the original Burges equation and used 
a third order term of  $C^3\cdot C_x$ .  
Later numerous generalization saw the light of sun by various authors, 
and the originality of the models are hard to identify.
The generalized Huxley equation \cite{genhux} has the 
source term of $\beta C(1-C^{\delta})(C^{\delta}-\gamma)$ 
where $\beta,\delta,\gamma$ are free real parameters.
 Lastly, we mention the  
the Burger's - Huxley and the  Burger's - Fisher equations \cite{burghux}. 
The first has the source term of $ - \gamma C_x 
+ \beta C(1-C^{\delta})(C^{\delta}-\gamma)   $ and the second of $- \gamma C_x + \beta C(1-C)$. 
Our presented list is of course far from being complete.  

Applications in different fields – for instance plasma physics or condensed matter – one may find in \cite{NePeCo2010,HeKa2005,PrDi2020}.   

We hope that this work may bring deeper understanding in the study of vapor diffusion 
\cite{ JANNOT2021121558, JIAO2021121543}, of the one dimensional convection- diffusion-reaction problem \cite{KENNEDY2003139, SENGUPTA2020109310, JAIN2021121465}, and of diffusive aspects in different flows \cite{Zhao-PhysRevE.66.036304}. 

Diffusion processes can be coupled to fluid dynamics phenomenon to describe the double (or multiple) diffusive convection phenomena \cite{mojt} where 
a convection is driven by two (or more) different density gradients described with different rates of diffusions.  \\ 
Another way of generalization of diffusion is the application of fractional derivatives.  
First consider when the time derivative is fractional. Such study was done in the work of Wyss \cite{wyss}.
The solutions were exactly given and could be expressed with the Fox functions. For the mathematical details of Fox functions discuss \cite{NIST}.  
About the space fractional diffusion equation one may find studies in \cite{YaBa2014}.  Comparison of our results to such functions could 
be the subject of a future study.  

 The analysis and control of coupled neural networks can be done with the reaction diffusion term as 
was given in the monograph of Wang {\it{et al}} \cite{wang}.  It is obvious, but we mention that diffusion equation has the same form as heat conduction 
equation. Its field has a mighty literature as well, from which we mention only two monographs \cite{akind,htrans}.   
Lastly, not to forget the field of continuously developing numerical methods of PDEs
 it is worthwhile to mention the new results obtained by  \cite{endre1,endre2}. 

\subsubsection{Self-similar Ansatz} 
We start the analysis with the self-similar Ansatz 
\begin{eqnarray}
C(x,t) = t^{-\alpha} f\left(\frac{x}{t^{\beta}} \right)  = t^{-\alpha} f(\eta),  
\label{ansatz}
\end{eqnarray}
where $\alpha$ and $\beta$ are the self-similar exponents being real numbers describing the decay and the spreading of the solution 
in time and space.  
These properties makes this Ansatz physically extraordinary relevant and was first introduced by 
Sedov \cite{sed}. For the present diffusion equation, after some trivial algebra we get: 
\eq 
\alpha =     \textrm{arbitrary real number},   \hspace*{3mm} \beta = 1/2,
\eqe
and there is a clear-cut time-independent ordinary differential equation (ODE) of 
\eq
-\alpha f - \frac{1}{2}\eta f' = D f''.  
\label{ode1}
\eqe
With the choice of $ \alpha = 1/2 $ and  setting the first integration constant to zero $c_1 = 0$ we get back the well-known Gaussian solution.  
 
This is the so-called fundamental solution and sometimes referred to as $ {\it{source \> type}}$ solution -- by 
mathematicians --  because for $t \rightarrow 0 $  the $C(x,0) \rightarrow  \delta(x)$.   
Otherwise with general $\alpha $, the solutions read as: 
\eq
f(\eta) = \eta  \cdot e^{-\frac{ \eta^2}{4D} }  \left(  c_1 M\left[1-\alpha  , \frac{3}{2} , \frac{ \eta^2}{4D} \right]  + c_2 U\left[1 -\alpha , \frac{3}{2} , \frac{ \eta^2}{4D} \right]    \right),
\label{f_eta2}
\eqe
 where  $M(\cdot,\cdot,\cdot) $ and $U(\cdot,\cdot,\cdot) $ are the Kummer's functions for details see the NIST Handbook \cite{NIST}. For non-negative integer $\alpha$s an alternative formulations of the result 
is possible in the form of    
\eq
f(\eta) = e^{-\frac{ \eta^2}{4D} }  \left(  \tilde{c}_1 H_{2\alpha-1}\left[\frac{\eta}{2\sqrt{D}}\right]  + 
\tilde{c}_2 \cdot \> _{1}F_{1}\left[\frac{1-2\alpha}{2} ,\frac{1}{2} ;\frac{ \eta^2}{4D} \right]    \right),
\label{f_etab}
\eqe 
where  $H_a(\eta)$ is the Hermite polynomial and  $ _{1}F_{1}(\cdot,\cdot;\cdot) $ is the hypergeometric function. 
The first part of the solution, the Hermite polynomials form a complete orthonormal 
basis set on the $-\infty.. + \infty $ range with the Gaussian weight function. 
(Note, that Hermite polynomials play an extraordinary role in quantum mechanics as the solution 
of the harmonic oscillator problem \cite{messiah} which pioneered the way to second quantization or field theory.)  

We think, that it is a remarkable result, in the sense that not only a special orthonormal function is 
obtain after solving the corresponding ODE but an extra not-orthogonal part. 
The $ _{1}F_{1}()  $ function with quadratic argument is not normalizable, such kind of functions sometimes arrised in our 
former investigations  \cite{imrelaci1, imrelaci2}.	 
From this point of view we may speak about a kind of "overcompleteness". Which we find unusual.   

\begin{figure}  
\scalebox{0.25}{
\rotatebox{0}{\includegraphics{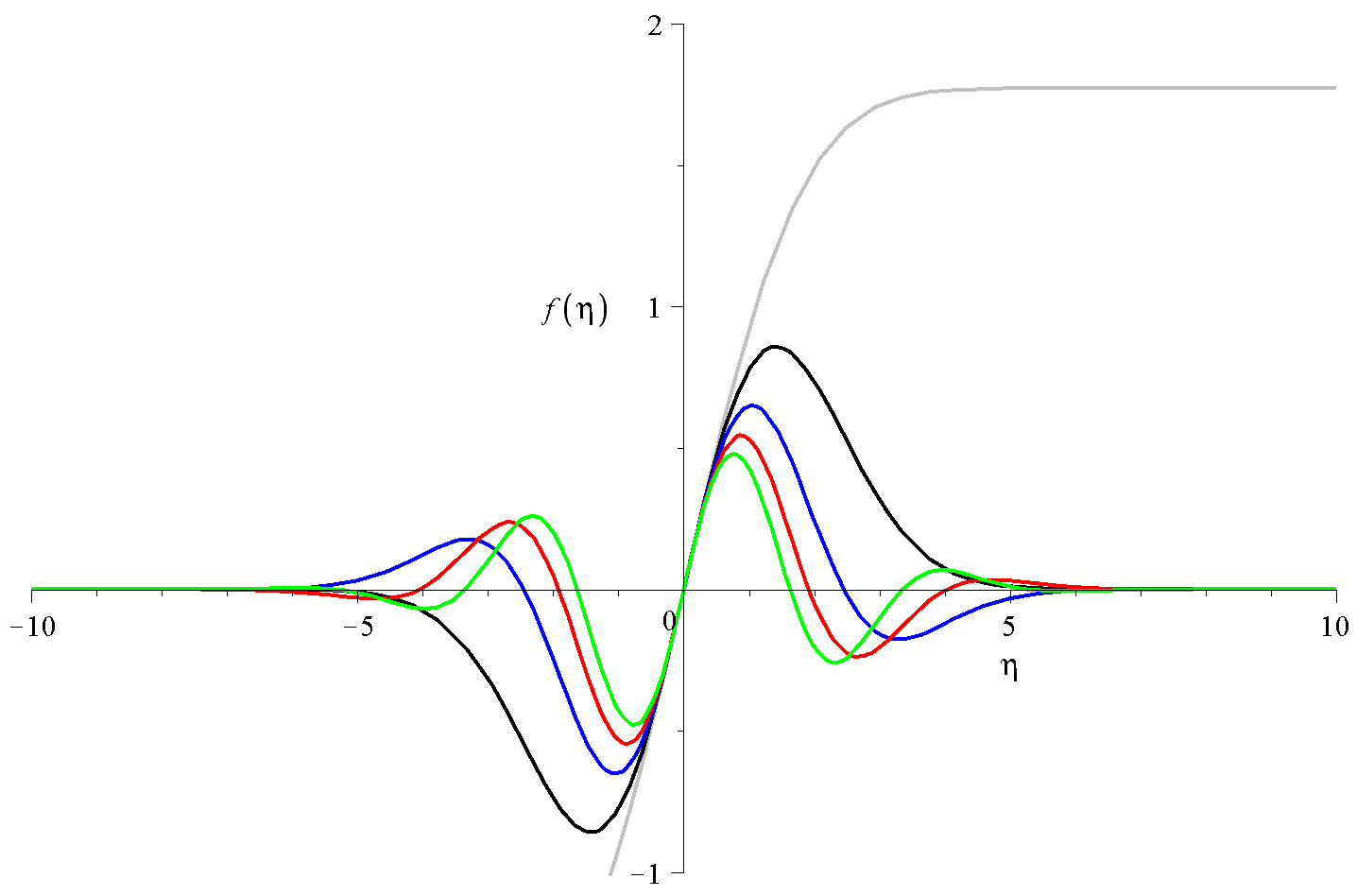}}}
\caption{Five evaluated shape functions $f(\eta)$ in Eq. \ref{fs}. The gray, black, blue, red and green curves 
are for $\alpha = 0,1,2,3$ and $4$, respectively. 
 Additional parameters $\kappa_1$ and $D$ are set to unity.	}
\label{egyes}      
\end{figure}

\begin{figure}[htp]
   \scalebox{0.35}{   {\includegraphics{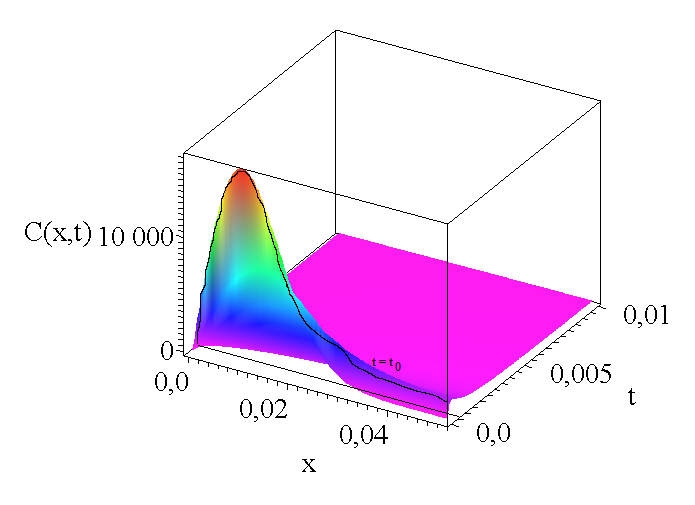}} } 
   \hspace*{0.8cm}    \scalebox{0.35}{   {\includegraphics{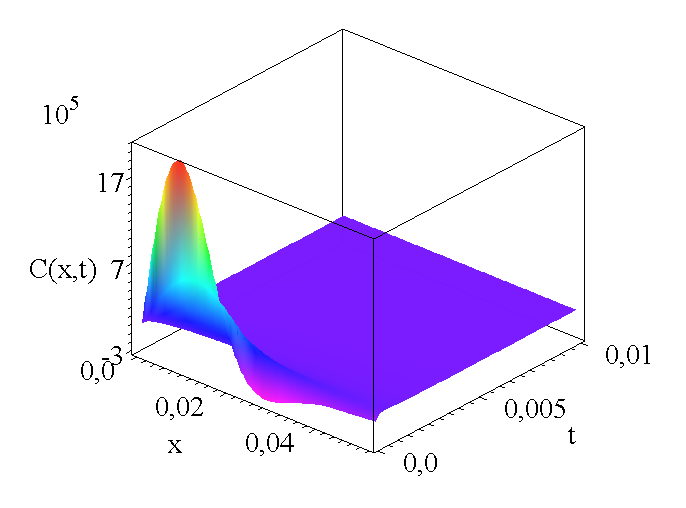}}}  \\
\hspace*{2cm}     
  $\alpha =  1 $     \hspace*{6cm} $\alpha = 3/2 $  \\ 
   \scalebox{0.4}{   {\includegraphics{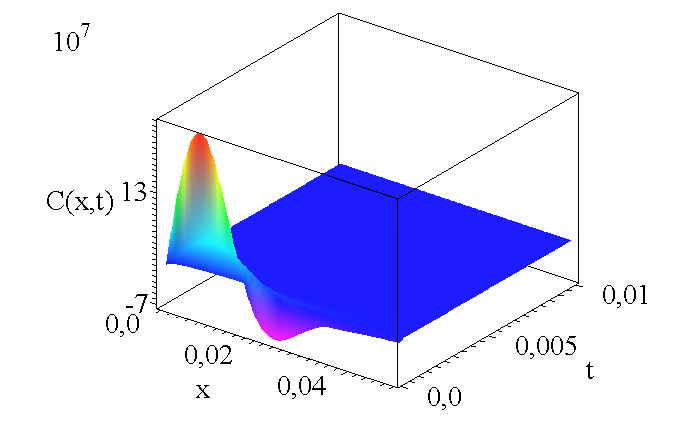}}} 
   \hspace*{0.8cm}    \scalebox{0.33}{   {\includegraphics{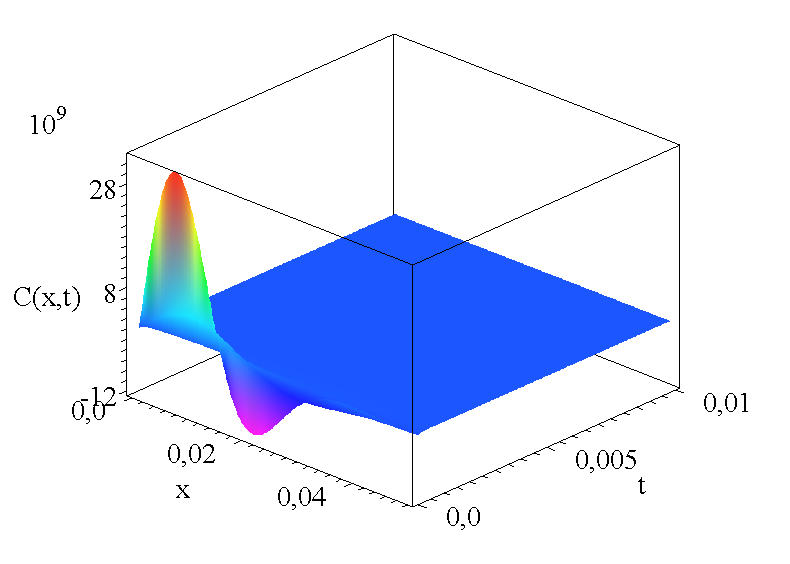}}} \\
    \hspace*{2cm}  $\alpha = 2 $     \hspace*{6cm} $\alpha = +\frac{5}{2}  $  \\
    \scalebox{0.35}{   {\includegraphics{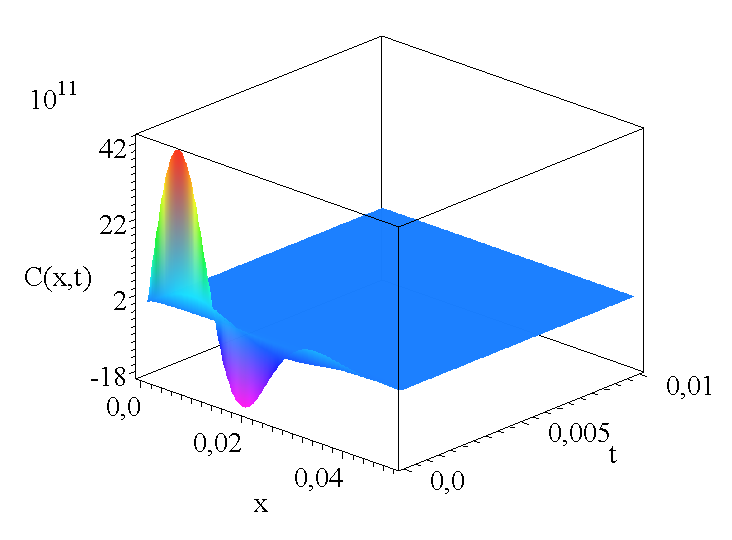}}} 
   \hspace*{0.8cm}    \scalebox{0.30}{   {\includegraphics{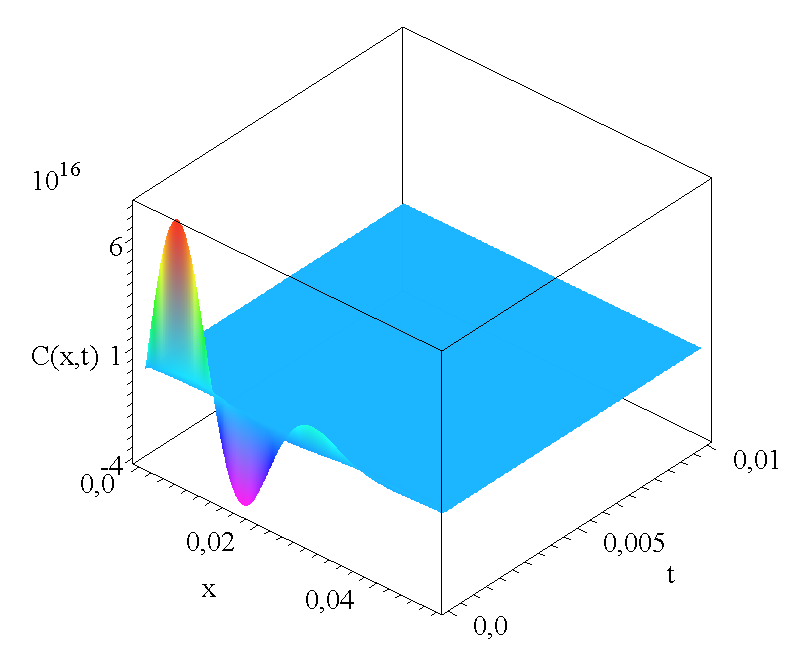}}} \\
    \hspace*{2cm}  $\alpha = 3 $     \hspace*{6cm} $\alpha = 4   $  \\
\caption{ The total solutions $C(x,t)$ with the shape function of Eq. (\ref{f_eta2}) for six various $\alpha$ values. 
Additional parameters $D = 2, c_2 = 1, c_2 = 0$ are the same 
in all cases. Note, that for a better comparison the same ranges are taken 
for the spatial and temporal variables in all six graphs. }  
\label{kettes}
\end{figure}

We present here the form of $f$  for the cases of $\alpha = 0,1,2,3,4$: 
\begin{eqnarray}
f(\eta) &=&      erf \left({\frac{\eta}{2\sqrt{ D}}} \right), \nonumber \\ 
f(\eta) &=& \kappa_0 \cdot \eta \cdot  e^{-\frac{\eta^2}{4D}},  \nonumber \\ 
f(\eta) &=&  \kappa_0 \cdot \eta  \cdot   e^{-\frac{\eta^2}{4D}}  \cdot \left( 1 -  \frac{1}{6D} \eta^2   \right), \nonumber  \\    
f(\eta) &=& \kappa_0 \cdot \eta \cdot e^{-\frac{\eta^2}{4D}} \cdot \left( 1 - \frac{1}{3D} \eta^2 
                         + \frac{1}{60} \frac{1}{D^2}  \eta^4 \right), \nonumber \\  
f(\eta) &=& \kappa_0 \cdot \eta \cdot e^{-\frac{\eta^2}{4D}} \cdot \left( 1 - \frac{1}{2D} \eta^2 
                         + \frac{1}{20} \frac{1}{D^2}  \eta^4 - \frac{1}{840} \frac{1}{D^3}  \eta^6   \right),
\label{fs}
\end{eqnarray}
the $\kappa_0$ is an arbitrary normalization constant. 
For completeness the final concentration distributions are also given,   
just inserting $\eta = x/t^{1/2}$ and the actual value of $\alpha$ 
we get: 
 \begin{eqnarray}
C(x,t) &=&      erf \left({\frac{x}{2\sqrt{ Dt}}} \right), \nonumber \\ 
C(x,t) &=&     \left( \frac{\kappa_1 x}{t^{\frac{3}{2}}}  \right)   e^{-\frac{x^2}{4Dt}}, \nonumber \\ 
C(x,t )&=& \left( \frac{\kappa_1 x}{t^{\frac{5}{2}}}  \right)    e^{-\frac{x^2}{4Dt}}  
 \left( 1 -  \frac{x^2}{6Dt}  \right), \nonumber  \\  
C(x,t) &=&     \left( \frac{\kappa_1 x}{t^{\frac{7}{2}}}  \right)  e^{-\frac{x^2}{4Dt}} \left( 1 - \frac{x^2}{3Dt}  
                         + \frac{x^4}{60(Dt)^2}    \right), \nonumber \\  
C(x,t) &=&    \left( \frac{\kappa_1 x}{t^{\frac{9}{2}}}  \right)    e^{-\frac{x^2}{4Dt}}  \left( 1 - \frac{x^2}{2Dt}  
                         + \frac{x^4}{20(Dt)^2}  - \frac{x^6}{840(Dt)^3}    \right).
\label{cs}
 \end{eqnarray}
Figure (\ref{egyes}) shows the given five shape functions. 
Functions with $\alpha > 0$ clearly show a decaying and oscillatory behavior. 
Figure(\ref{kettes}) shows six  $C(x,t)$s for different $\alpha$s, 
for generality we show two solutions for half-integer $\alpha$s as well.   
The quick decay and the slight oscillations are clear to see in all cases.  
Due to the linearity of the diffusion equation any linear combination of Eq. \ref{f_eta2} are automatically 
a solution as well enriching the possible mathematical structure of the diffusion process. 

\subsubsection{An interesting Ansatz} 
Just from curiosity we investigated the  "inverse self-similar Ansatz" in the form 
of: 
\eq
C(x,t) = x^{-\alpha} g\left(\frac{t}{x^{\beta}} \right)  = x^{-\alpha} g(\omega). 
\eqe
Now the role of the temporal and spatial variables is interchanged. 
The physical interpretation of this trial function is hard to see, till now 
we cannot find any kind of reasonable physical explanation of the results. 
(Just not to confuse with the original Ansatz we use the $g(\omega)$ notation for this case.)
After having done the usual derivation and algebraic steps we arrive at the  
relations of 
\eq 
\alpha =     \textrm{arbitrary real number},   \hspace*{3mm} \beta = 2.
\eqe 
The obtained ODE looks similar but obviously contains more terms then the 
previous ones
 \eq
4D \omega^2 g'' + \omega g'[-2\alpha D - 4D +2(\alpha-1)D] - g' + D\alpha(\alpha-1)g = 0.
\eqe
The solutions for the shape functions can be evaluated with the help of the usual Kummer's and exponential functions in the form of 
\begin{eqnarray}
 g &=  c_1 e^{-\frac{1}{4D\omega}}  \omega^{\left( \frac{5}{4} -  \frac{\sqrt{25-4\alpha^2 + 
4\alpha}}{4} \right)}   M \left( \frac{9}{4} +  \frac{\sqrt{25-4\alpha^2 + 4\alpha}}{4}  ,  
\frac{\sqrt{25-4\alpha^2 + 4\alpha}}{2} , \frac{1}{4D\omega}  \right)   +  \nonumber \\ 
& c_2 e^{-\frac{1}{4D\omega}} \omega^{\left( \frac{5}{4} -  \frac{\sqrt{25-4\alpha^2 + 4\alpha}}{4} \right)}
U \left( \frac{9}{4} +  \frac{\sqrt{25-4\alpha^2 + 4\alpha}}{4} , \frac{\sqrt{25-4\alpha^2 + 4\alpha}}{2}  , \frac{1}{4D\omega}  \right).
\label{inverz}
\end{eqnarray}
The two parameters of the Kummer's functions should be real therefore 
$\alpha$ must lie in the interval of $[\frac{1}{2} - \frac{\sqrt{26}}{2}, \frac{1}{2} - \frac{\sqrt{26}}{2} ]$ 
which is approximately $ -2.1 < \alpha < 3.1$ . 
Our experience showed, that basically for any numerical $\alpha$ values the shape functions have a 
power-law dependence like $f(\eta) \propto \eta^{n}$ where $  0 < n < 1$ and the $C(x,t)$s are 
divergent at large $x$ arguments. So we found no physically reasonable solutions, therefore 
present no figures for Eq. (\ref{inverz}).  
\subsubsection{A generalization} 
At this point it is straightforward to try the generalized form of the self-similar Ansatz  
\eq
C(x,t) = a(t) \cdot h\left( \frac{x}{b(t)} \right) =  a(t) \cdot h(\omega) , 
\label{ans22}
\eqe
where all $a,b$ and $h$ are continuous real functions with existing continuous 
first temporal and second spatial derivatives and $\omega$ is the new 
reduced independent variable.  
Note, that now the functions which are responsible for the time decay and spreading have a general form. 
Instead of the power law dependencies $t^{-\alpha}$   and $t^{\beta}$ we apply $a(t)$ and $b(t)$.   
Just calculating the needed temporal and spatial derivatives and plugging back to the original diffusion 
equation we arrive to the ODE of
\eq
a_t h -  \left( \frac{a b_t }{b} \right)  \omega h'  = \frac{Da}{b^2} h'',     
\label{third}
\eqe
where prime means derivation in respect to $\omega$ and subscript $t$ in respect to time.  
This equation should be an ODE for $h(\omega)$ therefore the coefficients of $h$ and $h'$ 
should be independent of time therefore should be equal to constants, so Eq. (\ref{third}) is became:
\eq
h -   \omega h'  = D h'',     
\label{third2}
\eqe
It has the solution of 
\eq
h(\omega) = 	C_1\omega +C_2\left(2De^{-\frac{\omega^2}{2D}}   
+ \sqrt{2\pi D}\omega\cdot  erf\left[ \frac{\sqrt{2}\omega}{2\sqrt{D}} \right] \right).
\eqe
The equations of constraints have to be fulfilled as well: 
\eq
b^2 a_t  = a\cdot b \cdot b_t = a, \hspace*{1cm} b^2 \ne 0.   
\eqe
The corresponding solutions can be easily obtained by direct integration and read:
\eq
a(t) = \pm \sqrt{2t + c_1},  \hspace*{1cm} 
b(t) = c_2\cdot  \sqrt{2t + c_1}.
\eqe 
Using the original definitions of the Ansatz (\ref{ans22}) we can obtain the 
final solution in the form of:
\begin{eqnarray}
C(x,t) =    (\pm \sqrt{2t + c_1})	\cdot \left( \frac{C_1 x}{ c_2\cdot  \sqrt{2t + c_1} } +C_2\left[2De^{-\frac{x^2}{2Dc_2^2(2Dt+c_1)}}   
 +  \right. \right. \nonumber \\ 
\left. \left.  \sqrt{2\pi D}\cdot \frac{x}{ c_2\cdot  \sqrt{2t + c_1} }  \cdot  erf\left\{ \frac{\sqrt{2 }x}{2c_2\sqrt{D(2t+c_1)}} \right\} \right] \right) .
\end{eqnarray}
Where $c_1,c_2,C_1,C_2$ are integration constants. Choosing $C_1 = c_1 = 0$ we get back the usual solution which is a sum of a Gaussian and and 
error function. 
Note, that this is equivalent to the self-similar solution where $\alpha = \beta = 1/2$. 
It is instructive to see that a more general form of the Ansatz does not necessarily lead to 
a larger class of solutions. The functions $a(t)$ and $b(t)$ do not have additional freedom, the power 
laws however have two free parameters -- $\alpha$ and $\beta$ two real numbers  -- which expand 
the class of possible solutions.  
Due to our best knowledge this relatively simple derivation is not yet published or widely known in the scientific community.  
At this point we have to note, that in our former study \cite{barna1} we investigated the traveling-profile Ansatz from 
\cite{behn} which interpolates between the traveling wave and the self-similar trial functions in the form of 
$C(x,t) = a(t)\cdot h([x-b(t)/c(t)] = a(f)h(\omega)$ where $a(t), b(t)$ and $c(t)$ are arbitrary continuous functions 
with existing first derivatives. The derived results were very similar to the Eq.   	$(\ref{f_eta2})$.   \\ 
For curiosity we checked the $C(x,t) = t^{-\alpha }f(\frac{x-ct}{t^{\beta}} )$ which 
is also an interpolation between the self-similar and the traveling wave Ansatz. 
Having done the spatial and temporal derivation we arrive at the usual constraint condition 
which dictates the proper values of the exponents. Unfortunately, we got now a contradiction 
to the exponent of $\beta$.  However, it became clear that the Ansatz can work well for 
first-order PDEs (like continuity or Euler equation) which we plan to take advantage of in our future investigations.  
\subsubsection{A redefinition of variables}
In the following we use the conjecture that the $f$ function can be written as: 
\begin{equation} 
f(\eta) = \eta e^{-\frac{\eta^2}{4D}} g(\eta).   
\end{equation}
It is worth to check this form to derive possible new results. 
The derivative of the function $f(\eta)$ is:
\begin{equation} 
f'(\eta) =  e^{-\frac{\eta^2}{4D}} g(\eta) - \eta \frac{\eta}{2D}  e^{- \frac{\eta^2}{4D}} 
                   + \eta e^{-\frac{\eta^2}{4D}} g’(\eta).    
\end{equation} 
The second derivative of function $f$ reads as follows 
\begin{eqnarray} 
&f''(\eta) = \nonumber \\ 
 &e^{-\frac{\eta^2}{4D}}  
     \left[  -\frac{-\eta}{2D} g(\eta) + g’(\eta) -\frac{-\eta}{D} g(\eta) +  \eta \frac{\eta^2}{4D^2} g(\eta) -       
                 \frac{\eta^2}{2D} g'(\eta) + g'(\eta)  \right.   \nonumber \\ 
& \left. -  \frac{\eta^2}{2D} g'(\eta) + \eta g''(\eta)  \right].
\end{eqnarray}
Inserting these functions into the equation, with having in mind that $\beta = \frac{1}{2}$ 
\begin{equation} 
-\alpha f - \frac{1}{2} \eta f' = D f'',  
 \end{equation}
we get for $g=g(\eta)$
\begin{equation} 
-\alpha \eta g = 2 D g’(\eta) - \eta g - \frac{\eta^2}{2} g' + \eta g''D.  
\end{equation}
Reordering the terms leads to 
\begin{equation}
\eta g'' + 2 g' -  \frac{\eta^2}{2D} g' + (\alpha-1) \frac{\eta}{D} g = 0.  
\end{equation}
The solutions read  
\begin{equation}
g(\eta) =  c_1 M\left[1-\alpha  , \frac{3}{2} , \frac{ \eta^2}{4D} \right]  + c_2 U\left[1 -\alpha , \frac{3}{2} , \frac{ \eta^2}{4D} \right],
\end{equation}
note, these are the same solutions as from the self-similar Ansatz, just in a separated form.  
These two examples clearly show that there is a relatively large freedom to define an 
Ansatz but only few of them lead to reasonable new solutions. 

\subsubsection{Using various series expansions of $f(\eta)$} 

As a possible generalization of the self-similar Ansatz 
we may define the following infinite power series of 
\eq 
C(x,t) = \sum_{i=1}^{\infty} a_i \cdot t^{-\alpha_i} \cdot  (f[\eta])^i,  
 \eqe
where $a_i$s and $\alpha_i$s are arbitrary real numbers. 
As first (and most logical case) just take the following two terms of: 
\begin{eqnarray}
C(x,t) = a t^{-\alpha} f\left(\frac{x}{t^{\beta}} \right)  + b t^{-\alpha} f\left(\frac{x}{t^{\beta}} \right)^2 = a t^{-\alpha} f(\eta) + b  t^{-\alpha} f(\eta)^2,   
\label{ansatz2}
\end{eqnarray}
the role of $\alpha$ and $\beta$ is still the same, and the role of $a$ and $b$ are to 
fix the ratio of the two components or -- which is more important -- to turn-on or turn-off one of them.   
It is clear that if we want to derive an ODE for the shape function the argument $\eta$ should remain on the 
same first power.  

After some trivial algebraic step we got the usual constraints for the two exponents 
\eq 
\alpha =     \textrm{arbitrary real number},   \hspace*{3mm} \beta = 1/2. 
\eqe
The derived  ODE is  
\eq 
-a\alpha f - \frac{a}{2}\eta f' - b(\alpha f^2  - \eta f f') = D(af'' +2b[f'^2 + ff"]),   
\label{ode11}\eqe 
for general $\alpha$ the solution is cumbersome containing large number of 
Kummer's M and Kummer's U functions and given in the Appendix at the end of the paper. 
However, for some given small values $\alpha = 1, \pm 1/2, \pm1, 3/2, 5/2$ the 
results can be expressed with the help of Gaussians and with the error function. 
The overall and complete function test of the solutions of Eq. (\ref{ode11}) is a hard question due to the five parameters of $\{D,a,b,c_1,c_2 \}$
($c_1$ and $c_2$ stand for the integration constants). \\ 
For $\alpha = 1/2$ and for arbitrary other real parameters the result reads the follows:  
\begin{eqnarray}
& f =  \nonumber \\ 
&-   \frac{ a e^{\frac{\eta^2}{4D}} \sqrt{-\frac{1}{D}} \pm   \sqrt{- e^{\frac{\eta^2}{4D}} \left[a^2 e^{\frac{\eta^2}{4D}} -4b c_1D \sqrt{- \frac{ \pi}{ D}} +   4Dbc_2\sqrt{-\frac{\pi}{D}}	erf \left\{ \frac{1}{2} \sqrt{\frac{-1}{D}} \eta \right\}  
 \right] /D} }{2 b e^{\frac{\eta^2}{4D}} \sqrt{-\frac{1}{D}}}, 
\label{sol1per2}
\end{eqnarray}
where erf is the usual error function. 
Note, that this solution let $(a = 0 \> \> \& \> \> b \> \epsilon  \> \mathbb{R})$ but not the opposite case. So 
it is impossible to get back the fundamental or Gaussian solution. 

\begin{figure}  
\scalebox{0.38}{
\rotatebox{0}{\includegraphics{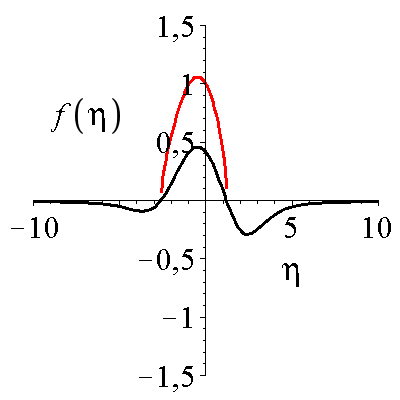}}} \hspace*{-0.8cm}  
\scalebox{0.35}{
\rotatebox{0}{\includegraphics{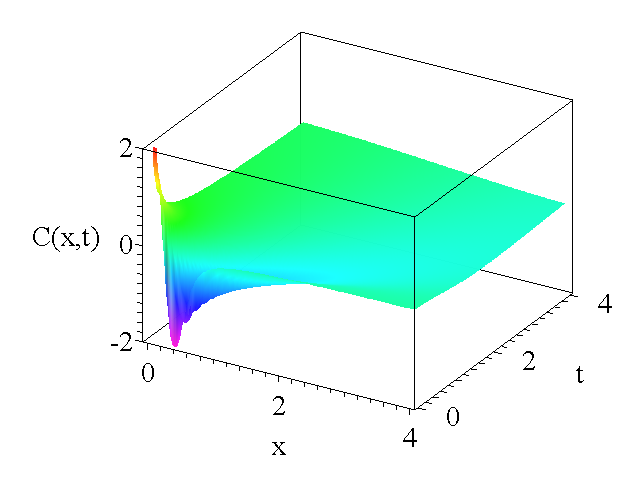}}}
\scalebox{0.28}{
\rotatebox{0}{\includegraphics{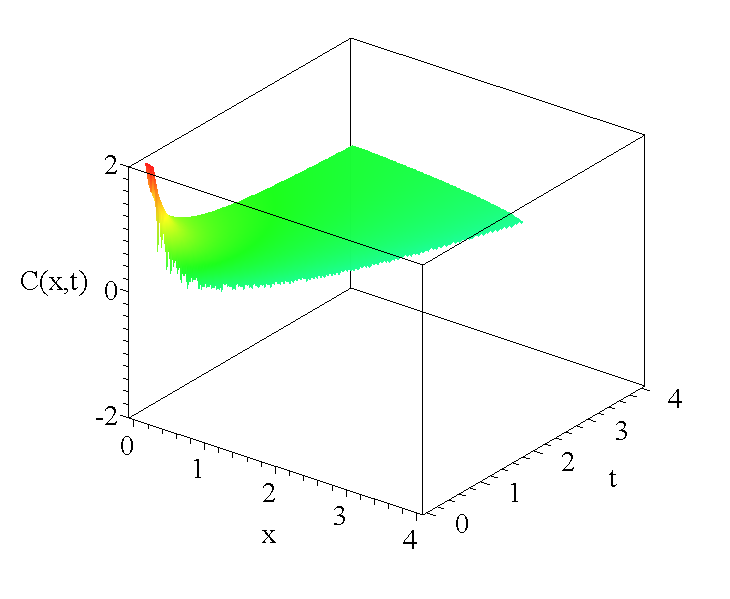}}}
\vspace*{-0.2cm} \\
\hspace*{1.4cm}  {\bf{a)}}   \hspace*{3cm}   {\bf{b)}} \hspace*{3cm}   {\bf{b)}}
\caption{ {\bf{a)}} Eq.  $(\ref{sol1per2})$  the black and red lines are for $a = 2, b = 1$ and for $a = 0,b = 1$ case for $D = c_1 = c_2 = 1$, 
 {\bf{b)}} the full solution $C(x,t)$ for  $a = 2, b = 1$ ,  
 {\bf{c)}} the full solution $C(x,t)$ for  $a = 0, b = 1$ .}
\label{egyes1}       
\end{figure}

Figure (\ref{egyes1}) shows the shape functions of  Eq. (\ref{sol1per2}) for  
$a = 2, b = 1$ and for $a = 0,b = 1$  and the corresponding final $C(x,t)$s 
as well. Note the remarkable new feature when the $f^2(\eta)$ term is considered alone, the solutions have a compact support. 
In this sense a linear PDE is reduced with a non-linear Ansatz to a non-linear ODE having non-linear properties.
Such a solution is definitely unknown for the scientific community.
We might begin to speculate about that even the regular diffusion equation could describe non-regular diffusion phenomena 
like the porous media equation \cite{velaz}. 
If we consider the quadratic term in (\ref{ansatz2}) only, we get a solution which is just  a bit  
simplified than the one given in the Appendix. Therefore we skip to present it.  \\ 
After this idea we may go a bit further, considering additional generalized forms like 
\eq 
C(x,t) =   t^{-\alpha} f\left(\frac{x}{t^{\beta}}\right)   +  \sum_{i=1}^{\infty} a_i \cdot t^{-\alpha_i} \cdot  \eta^i \cdot (f[\eta])^i, 
 \eqe
Just keeping the first two terms we arrive to 
\eq
C(x,t) =  t^{-\alpha} f\left(\frac{x}{t^{\beta}} \right)  +a t^{-\alpha} \eta  
f\left(\frac{x}{t^{\beta}} \right) = t^{-\alpha} f(\eta) + a  t^{-\alpha} \eta f(\eta),   
\label{ansatz3}
\eqe
which has the solutions of 
\eq
f(\eta) =\frac{   \left( e^{-\frac{ \eta^2}{4D} }  \left[  c_1 M\left\{1-\alpha  , \frac{3}{2} , 
\frac{ \eta^2}{4D} \right\}\eta  + c_2 U\left\{1 -\alpha , \frac{3}{2} , \frac{ \eta^2}{4D} \right\}\eta    \right] \right) }{a+b\eta}.   
\eqe
Note, that for $a,b > 0$ the only change is just a bit different scaling of the results. However if the $ a\cdot b < 0 $ the solution has an obvious singularity which makes it interesting but nonphysical which we ignore now. 
Last in this line we may try the Ansatz with the shape of:
\eq
C(x,t) =  at^{-\alpha} f\left(\frac{x}{t^{\beta}} \right)  +b t^{-\alpha}   f\left(\frac{x}{t^{\beta}} \right)^{1/2},  
\eqe
which unfortunately, gives no analytic solutions. 
Generally higher order terms in power expansions definitely gives higher degree non-linear  second-oder ODEs which hardly have analytic solutions. 

\subsubsection{Arbitrary self-similar exponents}  
 Last we arrived at a question which leads us out of the problem of  regular diffusion. 
We might ask -  if nothing else, only for the sake of completeness  -  what does it mean when 
both  $\alpha$, $\beta$  are arbitrary real numbers 
dictating the ODE of \eq
-  \alpha  f - \beta \eta f' =  D f''.  
\label{ode2}
\eqe
where D is still the usual diffusion coefficient. 
The solutions remain similar 
\eq
f(\eta) =   \eta e^{-\frac{\beta \eta^2}{2D} } \left(   c_1   M\left[\frac{2\beta -\alpha}{2\beta} , 
\frac{3}{2} , \frac{ \beta \eta^2}{2 D} \right]  + 
c_2  U\left[\frac{2\beta -\alpha}{2\beta} , \frac{3}{2} , \frac{ \beta \eta^2}{2 D} \right] \right) .   
\label{f_eta33}
\eqe 
With some easy reasoning we can find out the the original form of PDE.  An additional $ t^{2\beta - 1}$ time dependence 
has to be included to cancel the usual $\beta = 1/2$ constraint. 
So the starting PDE reads 
\eq
 \frac{\partial C(x,t)}{\partial t} = D \cdot  t^{2\beta -1} \cdot \frac{\partial^2 C(x,t) }{\partial x^2}.  \label{gen22} \\
\eqe
The role of the $\alpha$ parameter -- which is responsible for the temporal decay -- could be easily investigated as we could see above, 
however now the role of the $\beta$ -- which is responsible for the spreading -- was hidden till now. 
The solution functions are even functions therefore we just concentrate on 
positive arguments and investigate the Kummer's M function only. 
From the formula of (\ref{f_eta33})  two conditions are easy to notice. The first 
one is that $\beta \ne 0$ this is due to the denominator of the first parameter in 
the Kummer's function, and the second one is that for $ \beta < 0 $ the exponential 
multiplier function goes to infinity at large arguments. We ignore such kind of unphysical solutions. 

 The next figure \ref{alfa_beta}a)  presents various  $f(\eta)$ shape functions for $\alpha = 1$. 
Three cases can be distinguished:    \\
When $2\beta < \alpha$ the functions have zero transitions and show oscillatory behavior. \\    
When $2\beta = \alpha$ Kummer's function are equal to unity, hence the solution is purely Gaussian, with the quickest possible decay to zero.  \\ 
When $2\beta > \alpha$ the larger the beta the lower the global maximum and the slower the decay at large argument. 
The numerical value of $\alpha$ is irrelevant if it is positive. It is interesting that for negative $\alpha$s and for positive 
$\beta$s the total solution is again divergent at large arguments.  
For completeness we show   the $C(x,t)$ for $\alpha = 1$ and $\beta = 1$  on Fig. \ref{alfa_beta}b).  

\begin{figure} \hspace*{-1cm}  
\scalebox{0.5}{
\rotatebox{0}{\includegraphics{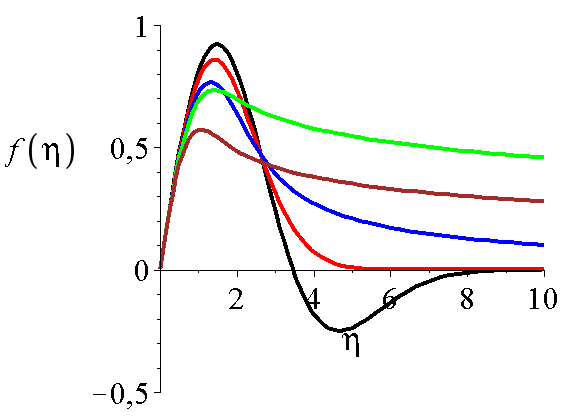}}}
\scalebox{0.4}{
\rotatebox{0}{\includegraphics{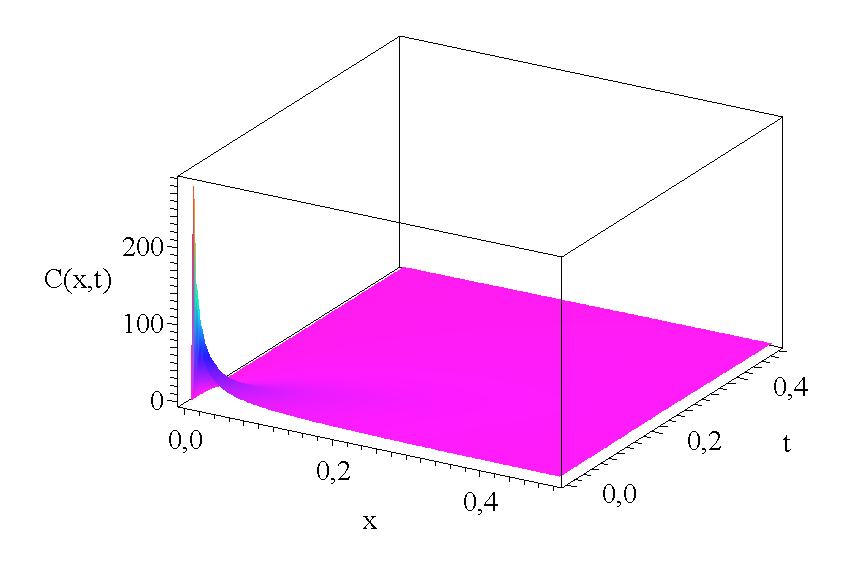}}}
\vspace*{-0.2cm} \\
\hspace*{1.4cm}  {\bf{a)}}   \hspace*{6cm}   {\bf{b)}} 
\caption{  {\bf{a)}} Functions of Eq. (\ref{f_eta33}) for $\alpha = 1$ where black,red,blue,green and brown lines are for 
$\beta = 1/4, 1/2, 1, 2$ and 3,  
{\bf{b)}}
The $C(x,t)$  solution  of Eq. (\ref{gen22}) for $\alpha = \beta = 1$.  }
\label{alfa_beta}       
\end{figure}
\begin{figure} 
\scalebox{0.65}{
\rotatebox{0}{\includegraphics{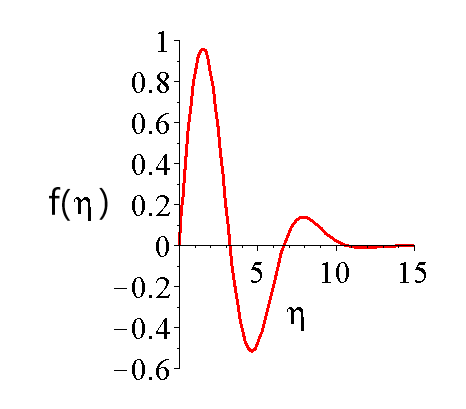}}}
\scalebox{0.36}{
\rotatebox{0}{\includegraphics{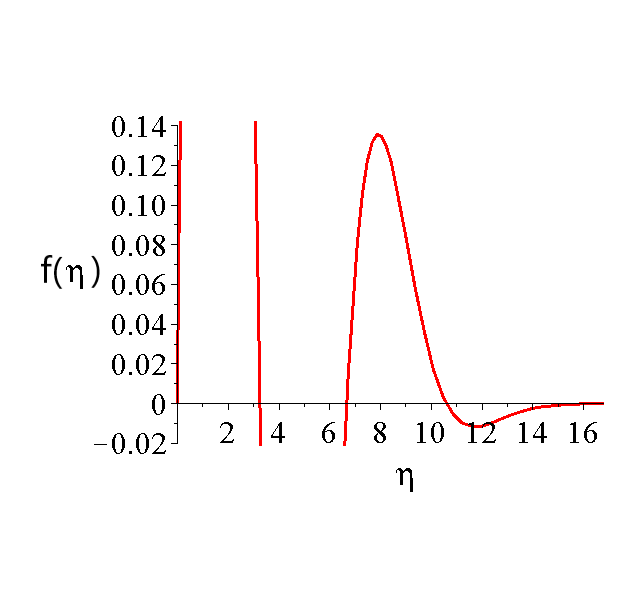}}}
\vspace*{-0.2cm} \\
\hspace*{1.4cm}  {\bf{a)}}   \hspace*{6cm}   {\bf{b)}} 
\caption{  {\bf{a)}} Functions of Eq. (\ref{f_eta33}) for $\alpha = 1$ when  
$\beta = 1/8$ ,  
{\bf{b)}}
magnification of $f(\eta)$ for values close to zero.  }
\label{beta=0-125}       
\end{figure}

   
One can see, that for sufficiently large values of $\beta$, the shape function $f$ has a maximum, which is followed by a relatively slow decay. 
For $\beta=1/4$ there is one root of the shape function $f(\eta)$, and correspondingly of the function $C(x,t)$. 
This shows that for values $\beta$ smaller than $1/2$ existence of nontrivial fluctuations in the value of $C(x,t)$ are possible, 
which means that it may become smaller than the average of the background.       
If $\beta$ is smaller than $1/4$ more roots of the 
shape function $f(\eta)$ are possible, on Fig. \ref{beta=0-125}. 
These values of $\beta$ shows a special behavior of the system, where the nontrivial diffusive effects may 
lead to temporal mass concentrations in case of mass diffusion.  
\section{Summary and Outlook}
We investigated the regular diffusion equation with the usual self-similar Ansatz and discussed all the solutions which are achievable beyond 
the Gaussian one. Most of the solutions can be described with the help of the 
Kummer's M and Kummer's U functions. For some special parameters 
the solutions go over the Hermite polynomials.  In the second part of the study we presented additional 
solutions which all can be derived from different modifications of the original self-similar Ansatz. Such solutions 
are again far from being well-known from the scientific public and therefor 
have to be published and discussed in details.  At the end of our manuscript  we investigated a special 
diffusion process which has time dependent diffusion coefficient.  Work is in progress to analyze spatial and temporal 
dependent diffusion equations which will be the topic of out next study. 
On the long run we would like to investigate reaction-diffusion equations (even with non-constant diffusion coefficients) as well.
Our experience so far suggests that numerous existing models can have new solutions with interesting features.     
\section{Appendix} 
To be complete we give the exact form of the solution of Eq.  (\ref{ode11}) for arbitrary $\alpha$. For a better 
transparency we introduce the following four abbreviation: 
\eq 
 M_{1-\alpha} := M\left(1-\alpha, \frac{3}{2}, \frac{\eta^2}{4D}\right), \hspace*{1cm} 
 U_{1-\alpha} :=  U\left(1-\alpha, \frac{3}{2}, \frac{\eta^2}{4D}\right), 
\eqe 
and similary 
\eq 
 M_{-\alpha} := M\left(-\alpha, \frac{3}{2}, \frac{\eta^2}{4D}\right), \hspace*{1cm} 
 U_{-\alpha} := U\left(-\alpha, \frac{3}{2}, \frac{\eta^2}{4D}\right).  
\eqe 
The solution formula is very elaborate but contains these four Kummer's functions only, therefore the notation 
is applicable.   
Note, that due to second degree of the ODE, two solutions exist: 
\begin{eqnarray}
&f(\eta) = -\frac{1}{2}\left(2aM_{1-\alpha}U_{-\alpha} + aM_{-\alpha}U_{1-\alpha} +2a\alpha M_{-\alpha}U_{1-\alpha}
\pm   \left[ 4a^2M_{1-\alpha}U_{-\alpha}^2    +  \right. \right. \nonumber \\ 
  &M_{-\alpha}M_{1-\alpha}U_{-\alpha}U_{1-\alpha}\{4a^2 +8a^2\alpha \}  + M_{-\alpha}U_{1-\alpha}\{ a^2 +4a^2\alpha \}   +
 \nonumber \\ 
& 4a^2\alpha^2M_{-\alpha}U_{1-\alpha} +  8c_2b \alpha M_{-\alpha}U_{1-\alpha}^2 - 
8c_1b\alpha M_{-\alpha}M_{1-\alpha}U_{1-\alpha} + \nonumber \\ 
  & 8c_2bM_{1-\alpha}U_{-\alpha}U_{1-\alpha}  -8c_1bM_{1-\alpha}U_{-\alpha} +  \nonumber \\ 
&\left. \left. 4c_2bM_{-\alpha}U_{1-\alpha} - 4bc_1 M_{-\alpha}M_{1-\alpha}U_{1-\alpha}
 \right]^{\frac{1}{2}}  \right)/  \nonumber \\ 
&  ( b [2\alpha M_{-\alpha} U_{1-\alpha} + 2M_{1-\alpha} U_{-\alpha} +
M_{-\alpha}  U_{1-\alpha}  ] ).  
\end{eqnarray}
\section{Acknowledgment}
One of us (I.F. Barna) was supported by the NKFIH, the Hungarian National Research 
Development and Innovation Office. 
\section{Conflicts of Interest}
 The authors declare no conflict of interest.
\section{Authors Contributions}
The corresponding author (Imre Ferenc Barna) had the original idea of the study, performed most 
of the calculations, created some of the figures and wrote large part of the manuscript. 
The second author (L\'aszl\'o M\'aty\'as) evaluated explicitly 
the form of the solutions for positive integer values of $\alpha$ and 
$\beta=1/2$, presented in the study, checked the literature of the investigated scientific field, improved 
the language of the final text and gave general instructions.  
\section{Data Availability}
The data that supports the findings of this study are all available within the article. 
\bibliography{mybibfile}
\end{document}